\documentstyle[twocolumn,prl,aps,epsf]{revtex}

\begin{document}

\preprint{UCTP-114-98}

\title{Dynamical Chiral Symmetry Breaking in QED
in a Magnetic Field: Toward Exact Results}

\author{V.P.~Gusynin$^{1}$, V.A.~Miransky$^{1,2}$,
and I.A.~Shovkovy$^{3}$\thanks{On leave of absence from
Bogolyubov Institute for Theoretical Physics, 252143,
Kiev, Ukraine.}}

\address{$^{1}$ Bogolyubov Institute for Theoretical Physics,
252143, Kiev, Ukraine\\
$^{2}$Department of Physics, Nagoya University,
Nagoya 464-8602, Japan \\
$^{3}$Physics Department, University of Cincinnati,
Cincinnati, Ohio 45221-0011}

\date{\today}
\draft

\maketitle

\begin{abstract}
We describe a (first, to the best of our knowledge) essentially
soluble example of dynamical symmetry breaking phenomenon in a 3+1
dimensional gauge theory without fundamental scalar fields:  QED in
a constant magnetic field.
\end{abstract}


Recently the magnetic catalysis of dynamical chiral symmetry
breaking has been established as a universal phenomenon in 2+1 and
3+1 dimensions: a constant magnetic field leads to the generation of
a fermion dynamical mass even at the weakest attractive interaction
between fermions \cite{1,2,3}. The essence of this effect is the 
dimensional reduction $D\to D-2$ in the dynamics of fermion
pairing in a magnetic field: at weak coupling, this 
dynamics is dominated by the lowest Landau level (LLL) which 
is essentially ($D-2$)-dimensional \cite{1,2,3}. The effect may 
have interesting applications in condensed matter physics 
\cite{cond-mat} and cosmology \cite{1,4,FI-cosm}.

In particular, this phenomenon was considered in 3+1 dimensional
QED \cite{2,3,4,5}. Since the dynamics of the LLL is
long-range (infrared), and the QED coupling constant is weak in the
infrared region, one may think that the rainbow (ladder)
approximation is reliable in this problem. As was shown in
Refs.~\cite{2,3,4}, the dynamical mass of fermions in this
approximation is
\begin{equation}
m_{dyn} = C \sqrt{|eB|}
\exp\left[-\frac{\pi}{2}
\left(\frac{\pi}{2\alpha}\right)^{1/2}
\right],
\label{m_dyn}
\end{equation}
where $B$ is a magnetic field, the constant $C$ is of order one and
$\alpha$ is the renormalized coupling constant related to the scale
$\mu^2\sim |eB|$.

Are higher order contributions indeed suppressed in this problem?
The answer is ``no". As was shown in Ref.~\cite{3}, because of the
(1+1)-dimensional form of the fermion propagator of the LLL
fermions, there are relevant higher order contributions. In
particular, considering this problem in the improved rainbow
approximation (when the vertex is bare, and the polarization
operator is calculated in one-loop approximation), it was shown
that, in all covariant gauges, the fermion mass $m_{dyn}$ is
given by Eq.~(\ref{m_dyn}) but with $\alpha\to \alpha/2$ \cite{3}.

As we wrote in the paper \cite{3}, ``it is a challenge to define the
class of all those diagrams in QED in a magnetic field that give a
relevant contribution in this problem". The aim of this letter is to
solve the problem. We will show that there exists a (non-covariant)
gauge in which the Schwinger-Dyson equations written in the improved
rainbow approximation are reliable. The expression for $m_{dyn}$
takes the following form:
\begin{equation}
m_{dyn} = \tilde C\sqrt{|eB|} F(\alpha)
\exp\left[-\frac{\pi}
{\alpha\ln\left(C_1/N\alpha\right)}
\right],
\label{m}
\end{equation}
where $N$ is the number of fermion flavors,
$F(\alpha)\simeq(N\alpha)^{1/3}$, $C_1\simeq 1.82\pm 0.06$ and
$\tilde C\sim O(1)$.
This expression for $m_{dyn}$ is essentially different from that in
the rainbow approximation (\ref{m_dyn}). As we will see, this
reflects rather rich and sophisticated dynamics in this problem.

The lagrangian density of massless QED in a magnetic field is
\begin{equation}
{\cal L}=-\frac{1}{4}F^{\mu\nu}F_{\mu\nu}+\frac{1}{2}
\left[\bar{\psi}, (i\gamma^{\mu}D_{\mu})\psi\right],
\label{lag}
\end{equation}
where the covariant derivative $D_{\mu}$ is
\begin{mathletters}
\begin{eqnarray}
D_{\mu}&=&\partial_{\mu}-i e (A_{\mu}^{ext}+A_{\mu}),\\
A_{\mu}^{ext}&=&\left(0,-\frac{B}{2}x_2,\frac{B}{2}x_1,0\right),
\label{vec-pot}
\end{eqnarray}
\end{mathletters}
i.e. we use the so called symmetric gauge for $A_{\mu}^{ext}$. The
magnetic field $B$ is in the $+x_3$ direction.

Besides the Dirac index (n), the fermion field carries an additional
flavor index $a=1,2,\dots,N$. Then the Lagrangian density in
Eq.~(\ref{lag}) is invariant under the chiral $SU_{L}(N)\times
SU_{R}(N)$ symmetry.

The Schwinger-Dyson (SD) equations in QED in external fields were
derived by Schwinger and Fradkin (for a review, see Ref.~\cite{6}).
The equation for the fermion propagator $G(x,y)$ is
\begin{eqnarray}
G(x,y)&=&S(x,y)-4\pi\alpha\int
d^4 u d^4 u^\prime d^4 z d^4 z^\prime
S(x,u)\gamma^\mu \nonumber \\
&\times& G(u,z) \Gamma^\nu (z,u^\prime,z^\prime)
G(u^\prime,y) {\cal D}_{\mu\nu}(z^\prime,u).
\label{SD-fer}
\end{eqnarray}
Here $S(x,y)$ is the bare fermion propagator in the external field
$A_{\mu}^{ext}$, and
${\cal D}_{\mu\nu}(x,y)$, $\Gamma^\nu (x,y,z)$ are the full photon
propagator and the full amputated vertex.

The full photon propagator satisfies the equations:
\begin{eqnarray}
{\cal D}^{-1}_{\mu\nu}(x,y)&=&D^{-1}_{\mu\nu}(x-y)
+\Pi_{\mu\nu}(x,y), \label{SD-pho} \\
\Pi_{\mu\nu}(x,y)&=&-4\pi\alpha \mbox{tr} \gamma_\mu
\int d^4 u d^4 z G(x,u) \nonumber \\
&&\times \Gamma_\nu (u,z,y) G(z,x),
\label{Pi_munu}
\end{eqnarray}
where $D_{\mu\nu}(x-y)$ is the free photon propagator and
$\Pi_{\mu\nu}(x,y)$ is the polarization operator.

The bare fermion propagator $S(x,y)$ in a constant
magnetic field was calculated
by Schwinger \cite{7}. In the symmetric gauge (\ref{vec-pot}), it
has the form:
\begin{equation}
S(x,y)=\exp\left(ie x^{\mu} A^{ext}_{\mu}(y)\right)
\tilde{S}(x-y).
\end{equation}
Then, it is not difficult to show directly from the SD equations
that
\begin{mathletters}
\begin{eqnarray}
&&G(x,y)=\exp\left(ie x^{\mu} A^{ext}_{\mu}(y)\right)
\tilde{G}(x-y),
\label{9a}\\
&&\Gamma(x,y,z)=\exp\left(ie x^{\mu} A^{ext}_{\mu}(y)\right)
\tilde{\Gamma}(x-z,y-z),
\label{9b}\\
&&{\cal D}_{\mu\nu}(x,y)=\tilde{{\cal D}}_{\mu\nu}(x-y),
\label{9c}\\
&&\Pi_{\mu\nu}(x,y)=\tilde{\Pi}_{\mu\nu}(x-y).
\label{9d}
\end{eqnarray}
\end{mathletters}
In other words, in a constant magnetic field, the Schwin\-ger phase
is universal for Green functions containing one fermion field, one
antifermion field, and any number of photon fields, and the full
photon propagator is translation invariant.

Our aim is to show that there exists a gauge in which the
approximation with a bare vertex,
\begin{equation}
\Gamma^{\mu}(x,y,z)=\gamma^{\mu} \delta(x-y)\delta(x-z),
\label{bare-ver}
\end{equation}
is reliable for the description of spontaneous chiral symmetry
breaking in a magnetic field.

We begin  by recalling the following facts concerning the problem of
the magnetic catalysis of chiral symmetry breaking \cite{1,2,3}:

1. At weak coupling, there is the LLL dominance in the dynamics of
fermion pairing. It is because of the presence of the large Landau
gap of order $\sqrt{|eB|}$, which is much larger than the dynamical
fermion mass $m_{dyn}$ (for weak coupling). In other words, higher
Landau levels decouple from the infrared dynamics with $k\ll
\sqrt{|eB|}$. This fact was explicitly
shown in the Nambu-Jona-Lasinio model \cite{3} and in QED \cite{8}.

2. The propagator $\tilde{S}(p)$ of fermions from the LLL is
\cite{2,3}
\begin{equation}
\tilde{S}(p)=2i e^{-(p_{\perp}l)^2}\frac{\hat{p}_{\parallel}+m}
{p^2_{\parallel}-m^2}O^{(-)},
\label{tildeS}
\end{equation}
where the magnetic length $l=|eB|^{-1/2}$, $p_{\perp}=(p^1,p^2)$,
$p_{\parallel}=(p^0,p^3)$, and
$\hat{p}_{\parallel}=p^0\gamma^0-p^3\gamma^3$. The operator
$O^{(-)}\equiv\left[1-i\gamma^{1}\gamma^{2} {\rm
sgn}(eB)\right]/2$ is the projection operator on the fermion
states with the spin polarized along the magnetic field. This point
and Eq.~(\ref{tildeS}) clearly reflect the (1+1)-dimensional
character of the dynamics of fermions in the LLL.

3. In the one-loop approximation, with fermions from the LLL, the
photon propagator takes the following form in covariant gauges
\cite{9,3}:
\begin{eqnarray}
{\cal D}_{\mu\nu}(k)&=&-i\Bigg[\frac{1}{k^2}g^{\perp}_{\mu\nu}
+\frac{k^{\parallel}_{\mu} k^{\parallel}_{\nu}}{k^2 k_{\parallel}^2}
+\frac{1}{k^2+k_{\parallel}^2 \Pi (k_{\perp}^2, k_{\parallel}^2)}
\nonumber\\
&\times&\left(g^{\parallel}_{\mu\nu}-\frac{k^{\parallel}_{\mu}
k^{\parallel}_{\nu}}{k_{\parallel}^2}\right)
-\frac{\lambda}{k^2}\frac{k_{\mu}k_{\nu}}{k^2}\Bigg],
\label{cov-gauge}
\end{eqnarray}
where $\lambda$ is a gauge parameter. The explicit expression for
$\Pi (k_{\perp}^2, k_{\parallel}^2)=\exp[-(k_{\perp} l)^2/2]
\Pi (k_{\parallel}^2)$ is given in Refs.~\cite{9,3}. For our
purposes, it is sufficient to know its asymptotes,
\begin{eqnarray}
\Pi (k_{\parallel}^2)\simeq \frac{\bar{\alpha}}{3\pi}
\frac{|eB|}{m^2} \quad \mbox{as}  \quad |k_{\parallel}^2| \ll m^2,
\label{Pi-IR}\\
\Pi (k_{\parallel}^2)\simeq -\frac{2\bar{\alpha}}{\pi}
\frac{|eB|}{k_{\parallel}^2} \quad \mbox{as}
\quad |k_{\parallel}^2| \gg m^2,
\label{Pi-UV}
\end{eqnarray}
where $\bar{\alpha}=N\alpha$. The polarization effects are absent
in the transverse components of ${\cal D}_{\mu\nu}(k)$. This is
because the bare vertex for fermions from the LLL is
$O^{(-)}\gamma^\mu O^{(-)}=O^{(-)}\gamma^\mu_\parallel$. Therefore
the LLL fermions couple only to the longitudinal $(0,3)$ components
of the photon field. Then, there is a strong screening effect in
the $\left(g^{\parallel}_{\mu\nu}-k^{\parallel}_{\mu}
k^{\parallel}_{\nu}/k_{\parallel}^2\right)$ component of the photon
propagator. For $m^2\ll |k_{\parallel}^2| \ll |eB|$ and 
$|k_{\perp}^2| \ll |eB|$, Eq.~(\ref{Pi-UV}) implies that
\begin{equation}
\frac{1}{k^2+k_{\parallel}^2 \Pi (k_{\perp}^2, k_{\parallel}^2)}
\simeq \frac{1}{k^2-M_{\gamma}^2},
\label{M_gamma}
\end{equation}
with $M_{\gamma}^2= 2\bar{\alpha}|eB|/\pi $.
This is reminiscent of the Higgs effect in the
(1+1)-dimensional QED (Schwinger model) \cite{10,11}.

We emphasize that infrared dynamics in this problem is very
different from that in the Schwinger model: since photon is
neutral, there is the four-dimensional
$k^2=k_{\parallel}^2-k_{\perp}^2$ in the denominator of the photon
propagator. However, the tensor and the spinor structure of this
dynamics is exactly the same as in the Schwinger model. This point
will be crucial for finding a gauge in which the improved rainbow
approximation [with the bare vertex (\ref{bare-ver})]  is
reliable\footnote{Since an external magnetic field does not lead to
confinement of fermions, their mass is gauge invariant in QED in a
magnetic field. Therefore {\it any} gauge can be used for the
calculations of the mass {\it if} either the calculations provide
the exact result or a good approximation is used, i.e., one can show
that corrections to the obtained result are small. Below we will
define such a gauge in this model.}.

We recall that, as was shown in Ref.~\cite{3}, despite
the smallness of $\alpha$, the expansion in $\alpha$ is broken in
covariant gauges in
this problem. The reason is that, because of the smallness of
$m_{dyn}$ in Eq.~(\ref{m_dyn}) as compared to $\sqrt{|eB|}$, there
are mass singularities, $\ln|eB|/m_{dyn}^2\sim \alpha^{-1/2}$, in
infrared dynamics. In particular, calculating the one-loop correction
to the vertex in covariant gauges with the photon propagator
(\ref{cov-gauge}), one finds that, when external momenta are of
order $m_{dyn}$ or less, there are contributions of order
$\alpha\ln^2(|eB|/m_{dyn}^2)\sim O(1)$. They come from the term
$k_{\mu}^{\parallel} k_{\nu}^{\parallel}/k^2 k_{\parallel}^2$ in
${\cal D}_{\mu\nu}(k)$ in Eq.~(\ref{cov-gauge}).

How can one avoid such mass singularities? A solution is suggested by
the Schwinger model. It is known that there is a gauge in which the
full vertex is just the bare one \cite{11}. It is the gauge with a
bare photon propagator
\begin{equation}
D_{\alpha\beta}(k)=-i\frac{1}{k^2}\left(g_{\alpha\beta}-\frac
{k_{\alpha} k_{\beta}}{k^2}\right)
-i d(k^2) \frac{k_{\alpha} k_{\beta}}{(k^2)^2},
\end{equation}
with the (non-local) gauge function $d= 1/(1+\Pi)$,
where the polarization function $\Pi(k^2)=-e^2/\pi k^2$ in the
Schwinger model (of course, here $\alpha,~\beta =0,1$). Then, the
full propagator is proportional to
$g_{\alpha\beta}$:
\begin{eqnarray}
{\cal D}_{\alpha\beta}(k)&=&D_{\alpha\beta}(k) +
i\left(g_{\alpha\beta}- \frac{k_{\alpha} k_{\beta}}{k^2} \right)
\frac{\Pi(k^2)}{k^2(1+\Pi(k^2))}
\nonumber \\
&=&-i \frac{g_{\alpha\beta}}{k^2(1+\Pi(k^2))}.
\label{gauge-S}
\end{eqnarray}
The point is that since now ${\cal D}_{\alpha\beta}(k)\sim
g_{\alpha\beta}$ and since the fermion mass $m=0$ in the Schwinger
model, all loop contributions to the vertex are proportional to
$P_{2n+1}\equiv  \gamma_{\alpha} \gamma_{\lambda_{1}} \dots
\gamma_{\lambda_{2n+1}} \gamma^{\alpha}=0$ in this gauge and,
therefore, disappear\footnote{$P_{2n+1}=0$ follows from the two
identities for the two-dimensional Dirac matrices: $\gamma_{\alpha}
\gamma_{\lambda} \gamma^{\alpha}=0$ and $\gamma_{\lambda_{i}}
\gamma_{\lambda_{i+1}}=g_{\lambda_{i} \lambda_{i+1}}
+\varepsilon_{\lambda_{i} \lambda_{i+1}} \gamma_{5}$
($\gamma_{5}=\gamma_{0} \gamma_{1}$, $\varepsilon_{\alpha\beta}
=-\varepsilon_{\beta\alpha}$, $\varepsilon_{01}=1$).}.

Let us return to the present problem. As it was emphasized above,
the tensor and the spinor structure of the LLL dynamics is
(1+1)-dimensional. Now, take the bare propagator
\begin{equation}
D_{\mu\nu}(k)=-i\frac{1}{k^2}\left(g_{\mu\nu}
-\frac{k_{\mu} k_{\nu}} {k^2}\right)
-i d(k_{\perp}^2, k_{\parallel}^2) \frac{k^{\parallel}_{\mu}
k^{\parallel}_{\nu}}{k^2 k_{\parallel}^2},
\end{equation}
with $d =-k_{\parallel}^2\Pi/[k^2+k_{\parallel}^2\Pi] +
k_{\parallel}^2/k^2$. Then, the
full propagator is
\begin{eqnarray}
&&{\cal D}_{\mu\nu}(k)=D_{\mu\nu}(k)
+i\left(g^{\parallel}_{\mu\nu}-
\frac{k^{\parallel}_{\mu} k^{\parallel}_{\nu}}{k_{\parallel}^2}
\right) \nonumber \\
&&\times
\frac{k_{\parallel}^2 \Pi(k_{\perp}^2, k_{\parallel}^2)}
{k^2[k^2+k_{\parallel}^2 \Pi(k_{\perp}^2, k_{\parallel}^2)]}
=-i\frac{g^{\parallel}_{\mu\nu}}{k^2+k_{\parallel}^2
\Pi(k_{\perp}^2, k_{\parallel}^2)} \nonumber\\
&&-i \frac{g^{\perp}_{\mu\nu}}{k^2}
-\frac{k^{\perp}_{\mu}k^{\perp}_{\nu} + k^{\perp}_{\mu}
k^{\parallel}_{\nu} + k^{\parallel}_{\mu}k^{\perp}_{\nu}}
{i(k^2)^2} .
\label{non-l}
\end{eqnarray}
The crucial point is that, as was pointed out above, the transverse
degrees of freedom decouple from the LLL dynamics. Therefore only
the first term in ${\cal D}_{\mu\nu}$, proportional to
$g^{\parallel}_{\mu\nu}$, is relevant.

Notice now that mass singularities in loop corrections to the
vertex might potentially occur only in the terms containing
$\hat{q}^{\parallel}_i=q_{i}^{0}\gamma^{0}-q_{i}^{3}\gamma^{3}$
from a numerator ($\hat{q}^{\parallel}_i+m_{dyn}$) of each fermion
propagator in a diagram (all other terms contain positive powers 
of $m_{dyn}$, coming from at least some  of the numerators and,
therefore, are harmless\footnote{For example, one can show that 
the contribution of the term with one-loop vertex correction in 
the SD equation is suppressed as $\alpha\ln\alpha$ with respect 
to the leading term.}). However, because of the same reasons as 
in the gauge (\ref{gauge-S}) in the Schwinger model, all those
potentially dangerous terms disappear in the gauge (\ref{non-l}).
Therefore all the loop corrections to the vertex are suppressed by
positive powers of $\alpha$ in this gauge. This in turn implies
that those loop corrections may result only in a change $\tilde
C\sim O(1)\rightarrow \tilde{C}^\prime\sim O(1)$  in Eq. (\ref{m}),
i.e., this expression yields the exact singularity at $\alpha=0$
for the fermion mass. In other words, in gauge (\ref{non-l}) there
exists a {\it consistent} truncation of the SD equations and the
problem is essentially soluble in this gauge\footnote{The 
gauge (\ref{non-l}) is unique in that. In other gauges,
there is an infinite set of diagrams giving relevant contributions
to the vertex. Therefore, in other gauges, one needs to sum up an
infinite set of diagrams to recover the same result for the fermion
mass.}.

As a result, in this gauge, the SD  equations (\ref{SD-fer}),
(\ref{SD-pho}) and (\ref{Pi_munu}) with the bare vertex
(\ref{bare-ver}) are reliable. They form a closed system of
integral equations. Using Eqs.~(\ref{9a})-(\ref{9d}) and the bare
propagator (\ref{tildeS}) of massless ($m=0$) fermions from the
LLL, one finds the SD equations for the full fermion propagator
\begin{equation}
\tilde{G}(p)=2i e^{-(p_{\perp}l)^2} \frac{ A(p_{\parallel}^2)
\hat{p}_{\parallel} +B(p_{\parallel}^2) }
{A^2(p_{\parallel}^2) p^2_{\parallel} - B^2(p_{\parallel}^2)}
O^{(-)}
\label{gen-sol}
\end{equation}
[compare with $\tilde{S}(p)$ in Eq.~(\ref{tildeS})]. In Euclidean
space they are: $A(p_{\parallel}^2)=1$ and
\begin{eqnarray}
B(p_{\parallel}^2)&=&\frac{\alpha}{2\pi^2} \int
\frac{d^2 q_{\parallel} B\left((p_{\parallel}
-q_{\parallel})^2\right)} {(p_{\parallel}-q_{\parallel})^2
+B^2\left((p_{\parallel} -q_{\parallel})^2\right)}
\nonumber\\
&&\times\int\limits_{0}^{\infty}\frac{dx \exp(-xl^2/2)}
{x+q_{\parallel}^2+q_{\parallel}^2\Pi_{E}(x,q_{\parallel}^2)},
\label{SD-B}
\end{eqnarray}
where the polarization function $\Pi$ is defined from
Eq.~(\ref{Pi_munu}) with a bare vertex.

A detailed analysis of these equations will be presented elsewhere.
Here we just indicate the crucial points in the analysis.

The polarization function $\Pi$ is a complicated functional of the
fermion mass function $B(p_{\parallel}^2)$. However, one can show
that the leading singularity, $1/\alpha\ln(\alpha)$, in
$\ln(m_{dyn}^2)$
in Eq.~(\ref{m}) is induced in the kinematic region with $m_{dyn}^2
\ll |q_{\parallel}^2| \ll |eB|$ and $m_{dyn}^2
\ll M_{\gamma}^2 \alt q_{\perp}^2 \ll |eB|$.
In that region, the fermions can be treated as massless, and
therefore the polarization function is $\Pi_{E}\simeq 2\bar{\alpha}
|eB| /\pi q_{\parallel}^2=M_{\gamma}^2/q_{\parallel}^2$ [see
Eqs.~(\ref{Pi-UV}) and (\ref{M_gamma})]. Therefore, in this
approximation, the photon propagator is a propagator of a free
massive boson with $M_{\gamma}^2=2\bar{\alpha}|eB|/\pi$.

\begin{figure}
\epsfxsize=6.5cm \epsfbox[55 510 340 750]{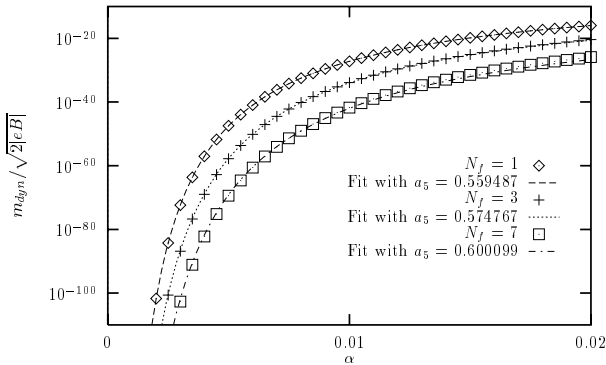}
\caption{Plot of the fit function and corresponding datapoints for
several values of $N_f$ when the only free parameter was $a_5$. }
\label{fig-fit}
\end{figure}

 The SD equation (\ref{SD-B}) with
$\Pi_{E}=M_\gamma^2/{q_{\parallel}^2}$ was solved both 
analytically and numerically\footnote{
The solution shows that the function $B(p_{\parallel}^2)$
is essentially constant for $p_{\parallel}^2\ll |eB|$,
$B(p_{\parallel}^2)=m_{dyn}$, and rapidly decreases for
$p_{\parallel}^2\gg |eB|$. Therefore this approximation is
self-consistent: the Ward identity for the vertex is satisfied in
the relevant kinematic region and the fermion pole is at
$p_{\parallel}^2=m_{dyn}^2$.}. In the numerical solution the 
following ansatz for $\ln (m_{dyn})$ was used,
\begin{equation}
\ln\frac{m_{dyn}}{\sqrt{2|eB|}}=\ln a_0 +{a_1\over3}\ln
\frac{N\alpha}{\pi}-\frac{a_2}{(\frac{\alpha}{\pi})^{a_3}\ln^{a_4}
\frac{a_5\pi}{N\alpha.}}.
\end{equation}
For small $\alpha$ ($0.001\leq\alpha\leq 0.1$) and different $N$
($1\leq N\leq7$) the best fit was found with
$a_0=a_1=a_2=a_3=a_4=1$ and $a_5\simeq 0.58\pm 0.02$ (see
Fig.~\ref{fig-fit}).
This fit corresponds to expression (\ref{m}) with $C_1=a_5\pi\simeq
1.82\pm 0.06$. The analytical solution yeilds a similar result.

The magnetic catalysis of chiral symmetry breaking in QED yields an
essentially soluble, and quite non-trivial, example of the
phenomenon of dynamical symmetry breaking in a (3+1)-dimensional
gauge theory without scalar fields. It may provide insight into the
non-perturbative dynamics of more complicated theories, such as
quantum chromodynamics. For example, chiral symmetry breaking in
this model is generated in the region of intermediate momenta, i.e.
it is independent of the deep infrared dynamics with $|q|\alt
m_{\rm dyn}$. It is noticeable as an example for a possibility
discussed in QCD: chiral symmetry breaking might be independent of
the dynamics of confinement with $|q|\alt\Lambda_{\rm QCD}\sim
m_{\rm dyn}$. Another noticeable point is the
dimensional reduction in the present model: there are arguments in
support of a similar reduction in the dynamics of chiral symmetry
breaking in QCD \cite{VafaWitten}.

{\bf Acknowledgments}.
We thank Anthony Hams for his generous help in numerical
solving the Schwinger-Dyson equation. The work of V.P.G. was
supported by Swiss  Grant No. CEEC/NIS/96-98/7 IP 051219. V.A.M.
acknowledges the JSPS (Japan Society for the Promotion of Science)
for its support during his stay at Nagoya University. He also
thanks K.~Yamawaki for his hospitality. The work of I.A.S. was
supported by U.S.  Department of Energy  Grant \#DE-FG02-84ER40153.


\begin{references}

\bibitem{1} V.~P.~Gusynin, V.~A.~Miransky, and I.~A.~Shovkovy,
\prl {\bf 73}, 3499 (1994); \prd {\bf 52}, 4718 (1995);
\pl B {\bf 349}, 477 (1995).

\bibitem{2} V.~P.~Gusynin, V.~A.~Miransky, and I.~A.~Shovkovy,
\prd {\bf 52}, 4747 (1995).

\bibitem{3} V.~P.~Gusynin, V.~A.~Miransky, and I.~A.~Shovkovy,
 Nucl.\ Phys.\ B {\bf 462}, 249 (1996).

\bibitem{cond-mat} K.~Farakos and N.~E.~Mavromatos, 
Int. J. Mod. Phys. B {\bf 12}, 809 (1998); 
G.~W.~Semenoff, I.~A.~Shovkovy and L.~C.~R.~Wijewardhana, 
Mod. Phys. Lett. A {\bf 13}, 1143 (1998).

\bibitem{4} C.~N.~Leung, Y.~J.~Ng, and A.~W.~Ackley, \prd
{\bf 54}, 4181 (1996); D.-S.~Lee, C.~N.~Leung, and Y.~J.~Ng,
\prd {\bf 55}, 6504 (1997); \prd {\bf 57}, 5224 (1998).

\bibitem{FI-cosm} E.~J.~Ferrer and V.~de~la~Incera, 
hep-ph/9810473.

\bibitem{5} D.~K.~Hong, Y.~Kim, and S.-J.~Sin, 
\prd {\bf 54}, 7879 (1996);
E.~J.~Ferrer and V.~de~la~Incera, \prd {\bf 58}, 065008 (1998);
V.~P.~Gusynin and A.~V.~Smilga, \pl B {\bf 450}, 267 (1999).

\bibitem{6} D.~M.~Gitman, E.~S.~Frad\-kin and  Sh.~M.~Shvarts\-man, 
in {\sl Quantum Electrodynamics with  Unstable Vacuum}, ed. by
V.~L.~Ginzburg (Nova Science, Commack,  NY, 1995).

\bibitem{7} J.~Schwinger, \prd {\bf 82}, 664 (1951).

\bibitem{8} D.~K.~Hong, \prd {\bf 57}, 3759 (1998).

\bibitem{9} Yu. M. Loskutov and V.~V.~Skobelev,
\pl A {\bf 56}, 151 (1976);
G.~Calucci and R.~Ragazzon, J.\ Phys.\ A {\bf 27},  2161 (1994).

\bibitem{10} J.~Schwinger, Phys.\ Rev. {\bf 125}, 397 (1962).

\bibitem{11} Y.~Frishman, in ``Particles, Quantum Fields
and Statistical Mechanics",
Lecture Notes in Physics, No. 32,
Ed. by M.~Alexanian and A.~Zepeda  (Springer, Berlin, 1975).

\bibitem{VafaWitten} G.~Vafa and E.~Witten, Nucl.Phys. B {\bf 234}, 173
(1984)

\end{references}
\end{document}